\documentclass[aps,prb,preprint]{revtex4}

\usepackage{graphicx}% Include figure files
\usepackage{dcolumn}% Align table columns on decimal point
\usepackage{bm}% bold math
\usepackage{amsmath}
\usepackage{amssymb}
\usepackage{footnpag}

\begin{document}
%\thispagestyle{empty}
%%%%%%%%%%%%%%%%%%%%%%%%%%%%%%%
\title{Electronic Transport in EuB$_{6}$}

\author{G.A. Wigger}
\author{R. Monnier}
\author{H. R. Ott}

\affiliation{Laboratorium f\"{u}r Festk\"{o}rperphysik,
ETH-H\"{o}nggerberg, CH-8093 Z\"{u}rich, Switzerland }

\author{D.P. Young}
   \altaffiliation{Present address: Department of Physics, Louisiana State University, Baton Rouge, LA 70803-4001 }
\author{Z. Fisk}

\affiliation{National High Magnetic Field Laboratory, Florida
State University, Tallahassee, Florida 32306}

\date{\today}

\begin{abstract}
EuB$_6$ is a magnetic semiconductor in which defects introduce
charge carriers into the conduction band with the Fermi energy
varying with temperature and magnetic field. We present
experimental and theoretical work on the electronic
magnetotransport in single-crystalline EuB$_6$. Magnetization,
magnetoresistance and Hall effect data were recorded at
temperatures between 2 and 300 K and in magnetic fields up to 5.5
T. The negative magnetoresistance is well reproduced by a model in
which the spin disorder scattering is reduced by the applied
magnetic field. The Hall effect can be separated into an ordinary
and an anomalous part. At 20 K the latter accounts for half of the
observed Hall voltage, and its importance decreases rapidly with
increasing temperature. As for Gd and its compounds, where the
rare-earth ion adopts the same Hund's rule ground state as
Eu$^{2+}$ in EuB$_{6}$, the standard antisymmetric scattering
mechanisms underestimate the $size$ of this contribution by
several orders of magnitude, while reproducing its $shape$ almost
perfectly. Well below the bulk ferromagnetic ordering at $T_C$ =
12.5 K, a two-band model successfully describes the
magnetotransport. Our description is consistent with published de
Haas van Alphen, optical reflectivity, angular-resolved
photoemission, and soft X-ray emission as well as absorption data,
but requires a new interpretation for the gap feature deduced from
the latter two experiments.
\end{abstract}
\maketitle
\section{Introduction}
The binary compound EuB$_6$ crystallizes in a simple cubic
lattice, with divalent Eu ions in their $^8$S$_{7/2}$ Hund's rule
ground state at the corners of the unit cell and B$_6$-octahedra
centered at the body-centered positions. With decreasing
temperature, it orders ferromagnetically \cite{matthias} via two
consecutive phase transitions at $\sim$ 15.5 K and $\sim$ 12.5 K,
respectively \cite {leo}, the first of which has recently been
interpreted as a phase separation between small magnetically
ordered regions with mobile charge carriers and large disordered
regions with localized magnetic polarons \cite{sullow}. A
spin-polarized electronic-structure calculation in the local
spin-density approximation for exchange and correlation (LSDA)
correctly reproduces the lattice constant, the internal
coordinates of the boron atoms and the size of the magnetic moment
in the stoichiometric compound \cite {massidda}. It also predicts
the system to be a semimetal, with overlapping conduction and
valence bands around the X-point of the Brillouin zone (BZ), in
contradiction with the results of a combined study based on
angle-resolved photoemission (ARPES) and bulk-sensitive soft X-ray
emission (SXE) and absorption (XAS) spectroscopies \cite
{denlinger}, which suggest a gap of at least 1 eV between the two
bands. While, given the approximation used in the calculation,
this discrepancy is not surprising, the fact that conduction band
states were observed at all in the photoemission experiment,
illustrates the fundamental problem one is faced with when trying
to describe the transport properties of this system and of the
hexaborides in general, namely that their behaviour is to a large
extent determined by defects and impurities \cite{aron,konrad}.

In this paper, we offer a quantitative analysis of magnetization,
magnetoresistance and Hall effect data, obtained on one and the
same sample, using a plausible model for the origin of the mobile
charge carriers and a consistent description of the dependence of
their concentration and their scattering rate on the applied
magnetic field and on temperature. Explicitely, we consider
EuB$_6$ to be a strongly compensated n-type magnetic
semiconductor. Due to the merging of defect states, i.e., boron
vacancy levels, with the conduction band, the latter acquires a
certain concentration of charge carriers, as evidenced by ARPES
\cite {denlinger}. In the paramagnetic state and in the absence of
external magnetic fields, these are equally distributed over six
pockets (three for each spin direction) centered at the X-points
of the BZ. Due to the thermal ionization of deep trap states in
the gap, the occupation of these states increases slightly with
temperature. The exchange coupling between the conduction
electrons and the localized (spin-) magnetic moments of the
Eu-ions leads to a lowering of the conduction band edge above
T$_c$ and to a splitting of the spin-up and spin-down bands below
T$_c$ \cite {balti, haas, massidda}. It is also responsible for
the so-called spin-disorder resistivity \cite {kasu, degennes,
dekker, kondo, haas} which, in a semiconductor, is strongly
dependent on the degree of spin-polarisation of the conduction
electrons and on the concommitant redistribution of the charge
carriers between the spin-up and spin-down bands \cite {haas}. We
have attempted to model our data on the anomalous Hall effect with
the mechanism suggested by Kondo \cite{kondo} for the case of
gadolinium metal, where the trivalent Gd ions adopt the same
4f$^7$ configuration as Eu$^{2+}$ in  EuB$_6$. We have generalized
this approach to include inelastic spin-flip processes. As in this
previous work, we find that the shape of the calculated field- and
temperature dependence of the anomalous Hall resistivity curves
matches that of the measured ones almost perfectly, but the
magnitude of the effect resulting from the calculation is several
orders of magnitude too small, if reasonable values for the
parameters are used in the theory. An alternative mechanism
proposed by Maranzana \cite{mara}, namely the interaction between
the orbital motion of the conduction electrons and the localized
moments, leads to the same functional dependence of $\rho_H$ on
temperature and applied field, but with an even smaller amplitude.

\section{Sample and Experimental Setup}

  The single-crystalline sample of EuB$_{6}$ was
  prepared by solution growth from Al flux. All measurements
  were made using the same platelet type specimen with dimensions of
  approximately 4.8 x 5.2 x 0.25 mm$^{3}$. The room temperature lattice
  constant of 4.185 \AA \ was evaluated from X-ray powder diffraction
  data, using a least-square refinement based on Cohen's method,
  with the software Xlat \cite{X-ray}. A Si spectrum served as the
  internal standard. Gold wires with 25 $\mu$m diameter were contacted to the sample with silver epoxy.
  All voltages were measured with a four-probe,
  low-frequency ac technique in the ohmic regime. The transverse
  magnetoresistance, which in the following is always referred to as
  the magnetoresistance, and the Hall voltage $V_{H}$ were measured
  in a configuration where the external field $\overrightarrow{B_{a}}$, between 0
  and 5.5 T, was oriented perpendicularly to both the applied current
  and the measured voltages, thus orthogonal to the platelet. The extended temperature range
  was covered by using a conventional $^{4}$He cryostat. Magnetization measurements were made
  in the same geometry with a commercial superconducting quantum interference device
  (SQUID) magnetometer, reaching temperatures between 2 and 330 K and
  magnetic fields up to 5.5 T.

\section{Electrical transport above 20 K}

\subsection{a.) Theory}

The unperturbed conduction electron levels are approximated by
parabolic bands, centered at the X-points of the BZ, with
spin-independent and, for convenience, isotropic effective masses
$m^*$. The energies of the band-bottoms are specified by
$\epsilon_{0}^{s}$, where $s = \pm$ labels the spin of the
electron. The corresponding Bloch states are denoted as $|$i
$\overrightarrow{k}$ s$\rangle$ , where i = 1, 2, 3 specifies the
X-point from which $\overrightarrow{k}$ is measured.

Following Haas \cite {haas}, we denote the eigenstates of the
system of magnetic Eu-ions by $|$$\alpha$$\rangle$ and their
occupation probabilities by $w_{\alpha}$, so that, for example,
the equilibrium value of the z-component of the spin of the ion
located at the site $\overrightarrow{R_n}$ is given by

\begin{equation}\label{equation1}
     \langle S_{nz} \rangle = \sum_{\alpha} w_{\alpha} \langle \alpha | S_{nz} | \alpha
\rangle \ \ \ .
\end{equation}

Our magnetization measurements show that above $\sim$ 20 K and up
to fields of 6 T, this quantity is well described by molecular
field theory and in the following, we shall assume that in this
temperature range, this also applies to higher-order correlation
functions.

The exchange interaction between the magnetic moments and the
mobile charge carriers has the form

\begin{equation}\label{equation2}
  \mathcal{H}_{1} = - \sum_{n=1}^{N}
  J(\overrightarrow{r}-\overrightarrow{R_{n}}) \overrightarrow{\mathbf{s}}\cdot\overrightarrow{S_{n}}
\end{equation}

where the sum is over all unit cells in the crystal, and
$\overrightarrow{\mathbf{s}}$ is the spin of the conduction
electron at $\overrightarrow{r}$. The range of
$J(\overrightarrow{r})$ is determined by the radius of the
4f-shell.

To first order in perturbation theory, this interaction produces
the following modification in the energy eigenvalues of the band
electrons

\begin{equation}\label{equation3}
  \begin{split}
  \Delta \epsilon_{s}^{(1)}(\overrightarrow{k}) &=  \sum_{\alpha}
  w_{\alpha} \langle i \overrightarrow{k} s; \alpha | \mathcal{H}_{1} | i \overrightarrow{k} s; \alpha
  \rangle \\
  &=  -\sum_{n=1}^{N} \langle i \overrightarrow{k}|
  J(\overrightarrow{r}-\overrightarrow{R_{n}}) | i
  \overrightarrow{k}\rangle \cdot \langle s | \overrightarrow{\mathbf{s}} |
  s \rangle \cdot \sum_{\alpha} w_{\alpha} \langle \alpha |
  \overrightarrow{S_{n}}| \alpha \rangle \ \ \ .
  \end{split}
\end{equation}

For a collinear ferromagnet, such as EuB$_6$,  the average value
embodied in the last sum points along the magnetization axis,
which we choose as the quantization axis for the conduction
electron spins. Furthermore, all magnetic ions being equivalent,
we can define

\begin{equation}\label{equation4}
     J_{i\overrightarrow{k}} = N \langle i \overrightarrow{k} |
     J(\overrightarrow{r})| i\overrightarrow{k}\rangle \approx J \ \ \
     ,
\end{equation}

where, in the last step, we have used the effective mass
approximation.

The spin-dependent energy shift then takes the form

\begin{equation}\label{equation5}
    \Delta \epsilon_{s}^{(1)} = - \frac{1}{2}s J \sigma  \ \ \ ,
\end{equation}

where the mean ionic spin $\sigma$=S(M/M$_{sat}$) depends on
temperature and magnetic field. M$_{sat}$=$-$(N/V)g$\mu$$_B$S is
the saturation magnetization per unit volume, V is the volume of
the sample, $\mu$$_B$ is the Bohr magneton, and $g$ is the
g-factor of the magnetic ion with spin S.

The second order correction is independent of the spin of the
conduction electrons in the absence of a net magnetization and is
completely dominated by the first order splitting (see eq.
(\ref{equation5})) otherwise, so that it may safely be neglected
in the discussion of magnetotransport properties.

The generic form for the matrix element of the interaction
$\mathcal{H}_2$ responsible for the antisymmetric scattering
between states with wave vectors $\overrightarrow{k}$ and
$\overrightarrow{k'}$ is \cite{kondo,mara}

\begin{equation}\label{H2}
\langle \overrightarrow{k'} \pm ; \alpha | \mathcal{H}_2 |
\overrightarrow{k} \pm ; \alpha \rangle = i
C\left(\theta_{\overrightarrow{k}\overrightarrow{k'}}\right)\left(\hat{k}\times\hat{k'}\right)
\cdot \hat{z} \sum_{n=1}^{N}
e^{i(\overrightarrow{k}-\overrightarrow{k'})\overrightarrow{R_n}}\langle
\alpha | S_{nz} | \alpha \rangle \ \ \ , \tag{6}
\end{equation}

where
$C\left(\theta_{\overrightarrow{k}\overrightarrow{k'}}\right)$ is
an even function of the scattering angle
$\theta_{\overrightarrow{k}\overrightarrow{k'}}$. This describes
the scattering along the direction perpendicular to
$\overrightarrow{k}$ and to the magnetization for spin-conserving
transitions, and does not affect the position of the energy bands
to first order in perturbation theory.

For the coupling between the orbital motion of the conduction
electrons and the localized spins we write, as proposed by
Maranzana, \cite{mara}

\begin{equation}\label{HamiltonMara}
  \mathcal{H}_2=\frac{\mu_0 e g \mu_B \hbar}{4\pi m^{*}}\cdot
  \sum_{n=1}^{N}\frac{\overrightarrow{S_n}\cdot\overrightarrow{L_n}}{|\overrightarrow{r}-\overrightarrow{R_n}|^3}
  \ \ , \ \ (e < 0, \mu_B > 0) \ \ ,
\tag{7}
\end{equation}

and

\begin{equation}\label{H11}
   C\left(\theta_{\overrightarrow{k}\overrightarrow{k'}}\right) =
   \frac{\mu_0 e g \mu_B \hbar}{m^{*}V}
   \frac{1}{cos^2\theta_{\overrightarrow{k}\overrightarrow{k'}}} \
   \ \ .
\tag{8}
\end{equation}

For the mechanism suggested by Kondo \cite{kondo} to explain the
anomalous Hall effect in gadolinium, which involves a virtual
excited state with one electron less in the 4f shell,

\begin{equation}\label{H13}
   C\left(\theta_{\overrightarrow{k}\overrightarrow{k'}}\right) =
   \frac{\lambda V_1^2}{2 S \Delta_{-}^2 N} \ \ \ ,
\tag{9}
\end{equation}

where $\lambda$ is the spin-orbit radial integral for 4-f
electrons in the 4f$^6$ configuration, $V_1$ is the mixing matrix
element between the $l$ = 1 component of the plane wave factor
$e^{i\overrightarrow{k}\overrightarrow{r}}$ of the Bloch function
(which is modulated by a function $u_{\overrightarrow{k}}$ with
the periodicity of the lattice and of pure d-character around each
Eu-site) and a 4f-orbital, and $\Delta_{-}$ is the minimum energy
necessary to excite one electron from the 4f$^7$ configuration to
the Fermi level.

The non-periodic part ($\mathcal{H'}$$_1$) of $\mathcal{H}$$_1$,
obtained by replacing the z-component of $\overrightarrow{S_n}$,
$S_{nz}$, by $S_{nz}-\sigma$ in eq. (\ref{equation2}), and
$\mathcal{H}$$_2$ induce transitions between different Bloch
states, the probability of which is given to lowest order by
Fermi's golden rule as

\begin{equation}\label{equation7}
   P^{(2)}(\overrightarrow{k}s;\alpha|\overrightarrow{k'}s';\alpha')
   =
   \frac{2\pi}{\hbar}\delta(\epsilon_{\overrightarrow{k}}^{s}-\epsilon_{\overrightarrow{k'}}^{s'}) |
   \langle \overrightarrow{k'}s';\alpha'
   |\mathcal{H}_{1}'+\mathcal{H}_2|\overrightarrow{k}s;\alpha\rangle|^{2}  \ \ \ .
\tag{10}
\end{equation}

The energy transfer between the conduction electrons and the
spin-system has been neglected (quasielastic or quasistatic
approximation), which is justified as long as the typical
excitation energy of the latter is smaller than the thermal energy
\cite {degennes}, as is certainly the case above $T_c$. The rhs of
eq. (\ref{equation7}) contains 4 terms. The two cross products
cancel due to the fact that the matrix elements of $\mathcal{H}_2$
are imaginary. From the definitions (\ref{H11}) and (\ref{H13}) it
is easy to see that the contribution from the square of the matrix
element of $\mathcal{H}_2$ is $a$ $priori$ irrelevant in the case
of Maranzana's mechanism and contributes less than one percent to
the total transition probability (\ref{equation7}) if a physically
reasonable value is used for the ratio $V_1/\Delta_{-}$ in Kondo's
model. Therefore we expect that the temperature and magnetic field
dependence of the resistivity is controlled entirely by the
exchange interaction between the conduction electrons and the
localized moments.

Two types of transition have to be considered, those without
spin-flip by

\begin{equation}\label{equation8a}
  \langle j \overrightarrow{k'}\pm; \alpha
  |\mathcal{H}_{1}^{'}|i\overrightarrow{k}\pm;
  \alpha\rangle = \mp
  \frac{1}{2N}\sum_{n=1}^{N}J_{\overrightarrow{k'}\overrightarrow{k}}^{ji}\cdot e^{[i(\overrightarrow{k}-\overrightarrow{k'}-\frac{\pi}{a}\hat{i}+\frac{\pi}{a}\hat{j})\cdot\overrightarrow{R_{n}}]}\langle\alpha|(S_{nz}-\sigma)|\alpha\rangle  \ \
  ,
  \tag{11a}
\end{equation}

and those with spin-flip by

\begin{equation}\label{equation8b}
  \langle j \overrightarrow{k'}\mp; \alpha\pm 1
  |\mathcal{H}_{1}^{'}|i\overrightarrow{k}\pm;
  \alpha\rangle = -
  \frac{1}{2N} \sum_{n=1}^{N}J_{\overrightarrow{k'}\overrightarrow{k}}^{ji}\cdot e^{[i(\overrightarrow{k}-\overrightarrow{k'}-\frac{\pi}{a}\hat{i}+\frac{\pi}{a}\hat{j})\cdot\overrightarrow{R_{n}}]}\langle\alpha\pm
  1|S_{n}^{\pm}|\alpha\rangle \ \ ,
  \tag{11b}
\end{equation}

where we have introduced the spin-raising and lowering operators
$S^{\pm }_{n}$ = $S_{nx}$ $\pm $ $iS_{ny}$ at the site
$\overrightarrow{R_n}$, and

\begin{equation}\label{equation8c}
   J_{\overrightarrow{k'}\overrightarrow{k}}^{ji}= N\cdot\langle j
   \overrightarrow{k'}|J(\overrightarrow{r})|i
   \overrightarrow{k}\rangle  \ \ \ .
   \tag{11c}
\end{equation}

For intravalley transitions ($i=j$), the momentum transfer is
small, and we can set $J^{ij}_{\bf kk'}$  equal to $J$ defined in
Eq. (\ref{equation4}) above. The short range of the exchange
integral in real space implies that the matrix element for
intervalley scattering will not be much reduced with respect to
$J$. Fortunately, the short wavelength of the associated spin
fluctuations and, in particular, the small range of scattering
angles available for this process, allow us to neglect it. The
same argument can be used to dismiss intervalley scattering in
(\ref{H2}).

To lowest order, the transition probabilities associated with the
matrix elements (\ref{equation8a}) and (\ref{equation8b}) are then
given by

\begin{equation}\label{equation9a}
   P^{(2)}(\overrightarrow{k}\pm;\alpha|\overrightarrow{k'}\pm;\alpha')
   = \frac{2\pi}{\hbar}\delta(\epsilon_{\overrightarrow{k}}^{\pm}-
   \epsilon_{\overrightarrow{k'}}^{\pm}) \left(\frac{J}{2N}\right)^2
   \cdot
   \sum_{n,n'}
   e^{[i(\overrightarrow{k}-\overrightarrow{k'})(\overrightarrow{R_{n}}-\overrightarrow{R_{n}'})]}
   \langle(S_{nz}-\sigma)(S_{n'z}-\sigma)\rangle  \ \ ,
   \tag{12a}
\end{equation}

and

\begin{equation}\label{equation9b}
   P^{(2)}(\overrightarrow{k}\pm;\alpha|\overrightarrow{k'}\mp;\alpha')
   = \frac{2\pi}{\hbar}\delta(\epsilon_{\overrightarrow{k}}^{\pm}-
   \epsilon_{\overrightarrow{k'}}^{\mp}) \left( \frac{J}{2N} \right) ^2
   \cdot
   \sum_{n,n'}
   e^{[i(\overrightarrow{k}-\overrightarrow{k'})(\overrightarrow{R_{n}}-\overrightarrow{R_{n}'})]}
   \langle S_{n}^{\pm} S_{n'}^{\mp} \rangle \ \ .
   \tag{12b}
\end{equation}

Following Haas \cite {haas}, we express the spin correlation
functions appearing in Eqs. (\ref{equation9a},\ref{equation9b}) in
terms of the generalized susceptibility per unit volume

\begin{equation}\label{equation10}
   \chi^{ij}(\overrightarrow{q}) =
   \frac{1}{V}\frac{(g\mu_{B})^2}{k_{B}T}\sum_{n,m}
   e^{[i\overrightarrow{q}\cdot(\overrightarrow{R_{n}}-\overrightarrow{R_{m}})]}
   \times \{\langle S_{ni}S_{mj}\rangle - \langle S_{ni}\rangle
   \langle S_{mj} \rangle \} \ \ \ ,
   \tag{13}
\end{equation}

where $i,j=x,y,z$. For a simple cubic (lattice constant $a$)
collinear ferromagnet, with one magnetic atom per unit cell,
$\chi^{ij}$ is diagonal and can be written as

\begin{equation}\label{equation11}
  \chi^{i}(\overrightarrow{q}) = [(\chi_{h}^{i})^{-1} + A
  q^2]^{-1}
  \tag{14}
\end{equation}

for small values of $\overrightarrow{q}$, where $\chi^{i}_{h}$ is
the susceptibility per unit volume of a single domain in a
homogeneous magnetic field and

\begin{equation}\label{equation12}
   A = \frac{Vk_{B}T_{C}a^{2}}{2N(g\mu_{B})^2S(S+1)}  \ \ \ .
   \tag{15}
\end{equation}

The transport relaxation rate for a Bloch state  $|
\overrightarrow{k} \pm\rangle$, with energy
$\epsilon^{\pm}_{\overrightarrow{k}}$ is then given by \cite
{haas}

\begin{equation}\label{equation13}
\begin{split}
   \frac{1}{\tau_{\overrightarrow{k}}^{\pm}}\equiv
   \frac{1}{\tau(\epsilon_{\overrightarrow{k}}^{\pm})} &= \frac{2\pi}{\hbar}k_{B}T \left(\frac{J}{2Ng\mu_{B}}\right)^{2}V\\
   &\times \sum_{\overrightarrow{k'}}[\chi^{z}(\overrightarrow{k'}-\overrightarrow{k})\delta(\epsilon_{\overrightarrow{k}}^{\pm}-\epsilon_{\overrightarrow{k'}}^{\pm})
   +2\chi^{x}(\overrightarrow{k'}-\overrightarrow{k})\delta(\epsilon_{\overrightarrow{k}}^{\pm}-\epsilon_{\overrightarrow{k'}}^{\mp})] \ \ \ .
\end{split}\tag{16}
\end{equation}

Inserting the explicit form of the susceptibilities and performing
the sum over $\overrightarrow{k'}$, we finally obtain, for spin
disorder scattering,

\begin{equation}\label{equation14}
\begin{split}
   \frac{1}{\tau(\epsilon_{\overrightarrow{k}}^{\pm})} = &
   \frac{1}{16\sqrt{2}\pi}\frac{\sqrt{m^{*}}k_{B}T}{\hbar^2\sqrt{\epsilon_{\overrightarrow{k}}^{\pm}}}
   \left(\frac{J}{Ng\mu_B}\right)^2 V^2
   \\ &\times\left[\frac{1}{A}ln\left(1+\frac{8m^{*}A}{\hbar^{2}}\chi_{h}^{z}\epsilon_{\overrightarrow{k}}^{\pm}
   \right)+\frac{2}{A}ln\left(\frac{1+\frac{2m^{*}A}{\hbar^{2}}\chi_{h}^{x}(\epsilon_{\overrightarrow{k}}^{\pm}+\epsilon_{\overrightarrow{k}}^{\mp}+2\sqrt{\epsilon_{\overrightarrow{k}}^{\pm}\epsilon_{\overrightarrow{k}}^{\mp}})}{1+\frac{2m^{*}A}{\hbar^{2}}\chi_{h}^{x}(\epsilon_{\overrightarrow{k}}^{\pm}+\epsilon_{\overrightarrow{k}}^{\mp}-2\sqrt{\epsilon_{\overrightarrow{k}}^{\pm}\epsilon_{\overrightarrow{k}}^{\mp}})}\right)\right] \ .
\end{split}
\tag{17}
\end{equation}

In the molecular-field approximation, we have \cite {haas}

\begin{equation}\label{equation15}
  \overrightarrow{M}=
  M_{sat}\hat{z}B_{S}(g\mu_{B}S|\overrightarrow{F}|/k_{B}T) \ \ \
  ,
  \tag{18}
\end{equation}

where $B_{S}$ is the Brillouin function for a spin $S$ acted upon
by the effective field

\begin{equation}\label{equation16}
  \overrightarrow{F}=\mu_{0}\gamma
  \overrightarrow{M}+\overrightarrow{B_{a}} \ \ \ .
  \tag{19}
\end{equation}

Here $\gamma$ is the molecular field constant and
$\overrightarrow{B_{a}}$ is the applied external field, so that

\begin{equation}\label{equation17}
   \chi_{h}^{x} = \chi_{h}^{y} =  M/B_{a} \ \ \ ,
   \tag{20}
\end{equation}

 and

\begin{equation}\label{equation18}
   \chi_{h}^{z} = \left[\left(M_{sat}\left(\partial B_{S} /
   \partial F\right)\right)^{-1}-\mu_0\gamma\right]^{-1} \ \ \ .
   \tag{21}
\end{equation}

The antisymmetric scattering responsible for the observed
anomalous Hall effect has its origin in the matrix element given
in eq. (\ref{H2}). For it to appear linearly in the transition
probability, we need to compute the latter to third order in the
matrix elements \cite {mara, kondo,irk}, which leads to

\begin{equation}\label{H3}
\begin{split}
   P^{(3)}&(\overrightarrow{k} \pm;\alpha|\overrightarrow{k'} \pm;\alpha)
   \approx \frac{2\pi}{\hbar}\delta(\epsilon_{\overrightarrow{k}}^{\pm}-
   \epsilon_{\overrightarrow{k'}}^{\pm}) \cdot \\ &\Re\left[\langle \overrightarrow{k'} \pm; \alpha | \mathcal{H}_{2} | \overrightarrow{k} \pm; \alpha
  \rangle \cdot \sum_{\overrightarrow{k''},s'',\alpha''}\frac{\langle \overrightarrow{k} \pm; \alpha| \mathcal{H}_{1} |\overrightarrow{k''} s''; \alpha '' \rangle\langle \overrightarrow{k''} s''; \alpha ''| \mathcal{H}_{1}|\overrightarrow{k'} \pm; \alpha \rangle
  }{\epsilon_{\overrightarrow{k}}^{\pm}-\epsilon_{\overrightarrow{k''}}^{s''}+i\delta}\right] \ \
  ,
\end{split}
\tag{22}
\end{equation}

where the real part is derived by use of the identity

\begin{equation}\label{H4}
   \frac{1}{\epsilon_{\overrightarrow{k}}^{s}-\epsilon_{\overrightarrow{k''}}^{s''}+i\delta}
   =
   \mathcal{P}\left(\frac{1}{\epsilon_{\overrightarrow{k}}^{s}-\epsilon_{\overrightarrow{k''}}^{s''}}\right)-i\pi
   \delta\left(\epsilon_{\overrightarrow{k}}^{s}-\epsilon_{\overrightarrow{k''}}^{s''}\right) \ \ \ .
\tag{23}
\end{equation}

In contrast to Kondo and Maranzana, we allow for spin-flip
exchange scattering to and from the summed-over intermediate
states, but the inelasticity in energy has again been neglected.
Inserting the appropriate matrix elements, we obtain

\begin{equation}\label{H5}
\begin{split}
   P^{(3)}(\overrightarrow{k} \pm;\alpha|\overrightarrow{k'} \pm;\alpha)
   = \frac{2\pi}{\hbar} \left(\frac{J}{2N}\right)^{2} C\left(\theta_{\overrightarrow{k}\overrightarrow{k'}}\right) &\delta(\epsilon_{\overrightarrow{k}}^{\pm}-
   \epsilon_{\overrightarrow{k'}}^{\pm}) (\hat{k} \times
   \hat{k'})\cdot
   \hat{z} \\ &\times
   \sum_{\overrightarrow{k''},s''}\delta(\epsilon_{\overrightarrow{k}}^{\pm}-
   \epsilon_{\overrightarrow{k''}}^{s''})\cdot
   D_{S}^{(3)}(\overrightarrow{k},\overrightarrow{k'},\overrightarrow{k''}) \ \ \
   ,
\end{split}
\tag{24}
\end{equation}

with

\begin{equation}\label{H6}
\begin{split}
   D_{S}^{(3)}(\overrightarrow{k},\overrightarrow{k'},\overrightarrow{k''})=
   \sum_{n,p,q} &
   e^{i\left[(\overrightarrow{k}-\overrightarrow{k'})\cdot\overrightarrow{R_{n}}+
   (\overrightarrow{k''}-\overrightarrow{k})\cdot
   \overrightarrow{R_{p}}+
   (\overrightarrow{k'}-\overrightarrow{k''})  \cdot
   \overrightarrow{R_{q}}\right]}\\ &\cdot \left[\langle
   S_{nz}(S_{pz}-\sigma)(S_{qz}-\sigma)\rangle +\langle
   S_{nz}S_{p}^{\mp}S_{q}^{\pm}\rangle \right] \ .
\end{split}
   \tag{25}
\end{equation}

In the spirit of molecular field theory, we now assume that the
sum over three-spin correlation functions can be limited to those
terms in which $n=p$ or $n=q$, and then make the following
decoupling

\begin{equation}\label{H7}
\begin{split}
   D_{S}^{(3)}(\overrightarrow{k},\overrightarrow{k'},\overrightarrow{k''})\simeq
   \sigma \sum_{p,q}&\left[
   e^{i(\overrightarrow{k''}-\overrightarrow{k'})\cdot(\overrightarrow{R_{p}}-
   \overrightarrow{R_{q}})}+ e^{i(\overrightarrow{k''}-\overrightarrow{k})\cdot(\overrightarrow{R_{p}}-
   \overrightarrow{R_{q}})}\right]\\ &\cdot \left[\langle
   (S_{pz}-\sigma)(S_{qz}-\sigma)\rangle +\langle
   S_{p}^{\mp}S_{q}^{\pm}\rangle \right]  ,
\end{split}
   \tag{26}
\end{equation}

which, with the definition of the generalized susceptibility
(\ref{equation10}), leads to

\begin{equation}\label{H8}
   D_{S}^{(3)}(\overrightarrow{k},\overrightarrow{k'},\overrightarrow{k''})
   = V \frac{k_{B}T}{(g\mu_{B})^2}\sigma\cdot
   \left[\chi^{z}(\overrightarrow{k''}\!-\!\overrightarrow{k'})+\chi^{z}(\overrightarrow{k''}-\overrightarrow{k})
   + 2 \chi^{x}(\overrightarrow{k''}-\overrightarrow{k'})+
   2\chi^{x}(\overrightarrow{k''}-\overrightarrow{k})\right] \ .
\tag{27}
\end{equation}

The sum over intermediate states in Eq. (\ref{H5}) is now
identical to the one appearing in the expression for the transport
relaxation rate of Eq. (\ref{equation13}), with
$\epsilon_{\overrightarrow{k''}}^{s''}$ =
$\epsilon_{\overrightarrow{k''}}^{\pm}$ or
$\epsilon_{\overrightarrow{k''}}^{\mp}$ for spin conserving or
spin-flip transitions, respectively. This allows us to write the
transition probability for skew scattering in the compact form

\begin{equation}\label{H9}
   P^{(3)}(\overrightarrow{k}\pm;\alpha | \overrightarrow{k'} \pm;\alpha)
   = 2 C\left(\theta_{\overrightarrow{k}\overrightarrow{k'}}\right) \delta(\epsilon_{\overrightarrow{k}}^{\pm}-
   \epsilon_{\overrightarrow{k'}}^{\pm})\cdot {\left(\hat{k}\times\hat{k'}\right)\cdot
   \hat{z}}\cdot
   \frac{1}{\tau(\epsilon_{\overrightarrow{k}}^{\pm})}\cdot \sigma \ \
   ,
\tag{28}
\end{equation}

where we have used the fact that the relaxation rate of a
particular Bloch state only depends on its energy. Given an
interaction with the above angular dependence, we can use the
exact result derived by Fert \cite{fert} for the Hall resistivity
$\rho_{H}$ (his Eq. (15)), to define the relaxation rate for
antisymmetric scattering

\begin{equation}\label{H10}
   \frac{1}{\tau_{as}(\epsilon_{\overrightarrow{k}}^{\pm})}=
   \frac{V}{8\pi^{3}}\int d^{3}k' P^{(3)}(\overrightarrow{k}\pm;\alpha|\overrightarrow{k'}
   \pm;\alpha)\hat{k'}\cdot \hat{y} = \widetilde{C}\cdot k \frac{1}{\tau(\epsilon_{\overrightarrow{k}}^{\pm})}\cdot
   \sigma \ \ \ ,
\tag{29}
\end{equation}

where, for the interaction proposed by Maranzana \cite{mara}

\begin{equation}\label{H12}
   \widetilde{C}^{M} = \frac{\mu_0 e g \mu_B}{2 \pi^2 \hbar} = 1.794\cdot10^{-15}
   \ \textrm{m} \ \ \ .
\tag{30}
\end{equation}

For the mechanism, suggested by Kondo \cite{kondo} to explain the
anomalous Hall effect in gadolinium metal,

\begin{equation}\label{H14}
   \widetilde{C}^{K} =
   \frac{m^{*} \Omega_{cell}\lambda V_1^2}{6 \pi^2 S \Delta_{-}^{2}\hbar^2}
   \ \ \ ,
\tag{31}
\end{equation}

where $\Omega_{cell}$ is the volume of the unit cell.

In our comparison with experiment, the constant $\widetilde{C}$ in
eq. (\ref{H10}) will be treated as a free parameter.

\subsection{b.) Experimental Results and Analysis}

\subsubsection{Temperature dependent electrical
resistivity in zero magnetic field}\label{rhoTT}

The temperature dependent electrical resistivity arises from
several scattering processes, which we shall, as usual, consider
as independent (Matthiessen's rule). From the theory developed
above we can calculate the conductivity of the coupled spin-up and
spin-down charge carriers in the presence of spin-disorder
scattering only according to

\begin{equation}\label{equation28}
  \sigma_{sd}^{\pm}(B_{a},T) = -\frac{2e^2}{3m^*}\int_{-\infty}^\infty
  \tau_{sd}^{\pm}(\epsilon)(\epsilon-\epsilon_{0}^{\pm})g^{\pm}(\epsilon)\frac{\partial f_{0}(\epsilon)}{\partial
  \epsilon} d\epsilon
  \tag{32}
\end{equation}

with the Fermi function

\begin{equation}\label{equation29}
  f_{0}(\epsilon) =
  \frac{1}{1+exp(\frac{\epsilon-\zeta}{k_{B}T})} \ \ ,
  \tag{33}
\end{equation}

where $\zeta$ stands for the chemical potential. The density of
states for each spin orientation is

\begin{equation}\label{equation30}
  g^{\pm}(\epsilon)=\frac{3}{4\pi^2}\left(\frac{2m^*}{\hbar^2}\right)^{3/2}(\epsilon-\epsilon_{0}^{\pm})^{1/2}
  \ \ \ .
  \tag{34}
\end{equation}

The relaxation times $\tau_{sd}^{\pm}(\epsilon) \equiv
\tau(\epsilon_{\overrightarrow{k}}^{\pm})$ are given by Eq.
(\ref{equation14}). In the absence of a magnetic field and in the
temperature range considered here, the system is unpolarized
($\epsilon_{0}^{+}=\epsilon_{0}^{-}$), and the relaxation time is
the same for both spin orientations. We can then define the
spin-disorder resistivity as

\begin{equation}\label{equation31}
   \rho_{sd}(0,T) =
   \left(\sigma_{sd}^+(0,T)+\sigma_{sd}^-(0,T)\right)^{-1} \ \ \ .
   \tag{35}
\end{equation}

In order to calculate the contribution $\rho_{ph}$ of the
electron-phonon interaction to the resistivity, we use the model
that was recently suggested by Mandrus and collaborators
\cite{mand} for LaB$_{6}$. The electrons are assumed to be
scattered by localized low-frequency Einstein oscillators,
corresponding to the almost independent motion of the rare-earth
ions in their boron "cages", as well as by Debye-type phonons due
to the collective motion of the boron framework. In a first step,
we apply the model, described in detail in ref. \cite{mand}, to
fit the resistivity data of YbB$_{6}$ \cite{chiao}. In this
compound, the Yb cations also adopt a divalent configuration but
they carry no magnetic moment. We then renormalize the obtained
Einstein frequency by the square root of the mass ratio between Yb
and Eu, leading to $\theta_{E}$=168 K for EuB$_6$. The Debye
frequency ($\theta_{D}$=1160 K), to which the results are not
sensitive to start with, is left unchanged. Next, we have to
account for the resistivity $\rho_{d}$ arising from the scattering
of the conduction electrons at point defects. We anticipate the
charge carrier density to be high enough to efficiently screen the
latter and therefore, the corresponding relaxation rate can be
considered as temperature-independent. The total resistivity is
then given by

\begin{equation}\label{equation32}
   \rho(0,T) = \rho_{sd}(0,T)+\rho_{ph}(0,T)+\rho_{d}(0,T) + \rho_{cont} \ \ \
   ,
   \tag{36}
\end{equation}

where $\rho_{cont}$ is a (small) contribution arising from
non-ideal electrical contacts to the sample, and which we assume
to be independent of temperature and magnetic field.

\begin{figure}
  \centering
  \includegraphics[width=\linewidth]{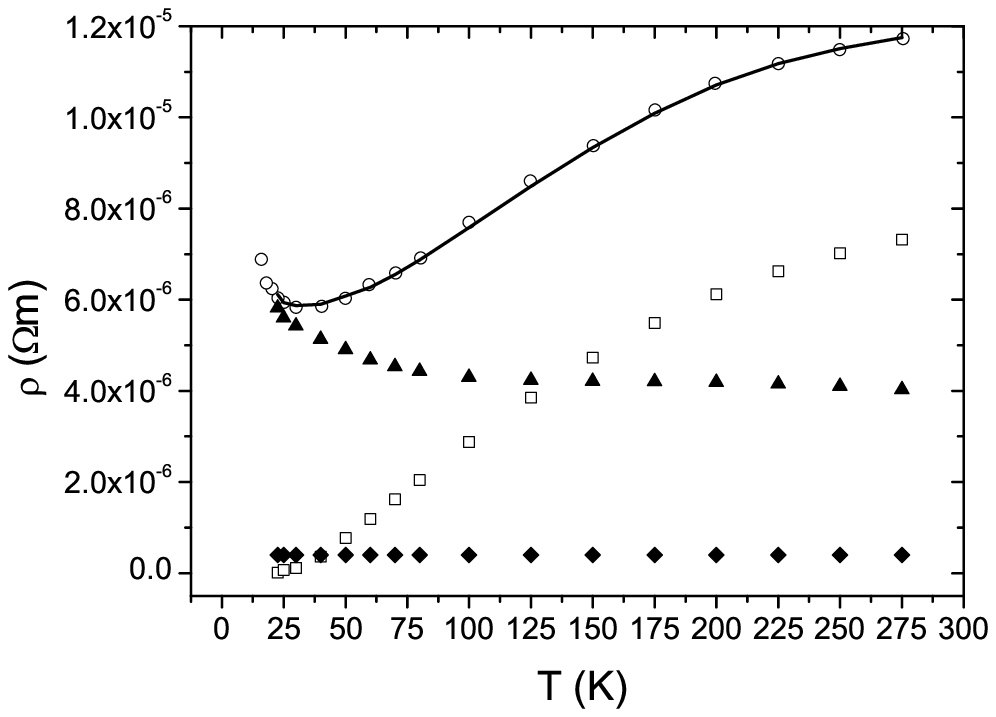}
  \caption{The total measured resistivity $\rho$ of EuB$_6$ is
represented by open circles. The calculated contributions due to
scattering by phonons and magnetic excitations are shown by open
squares and closed triangles, respectively. The closed diamonds
represent the combined contribution to the resistivity of the
scattering by point defects and of the non-ideal contacts. The
solid line represents the total calculated resistivity in zero
external field above 20 K.}
 \label{rhovsT}
\end{figure}

\begin{figure}
  \centering
  \includegraphics[width=\linewidth]{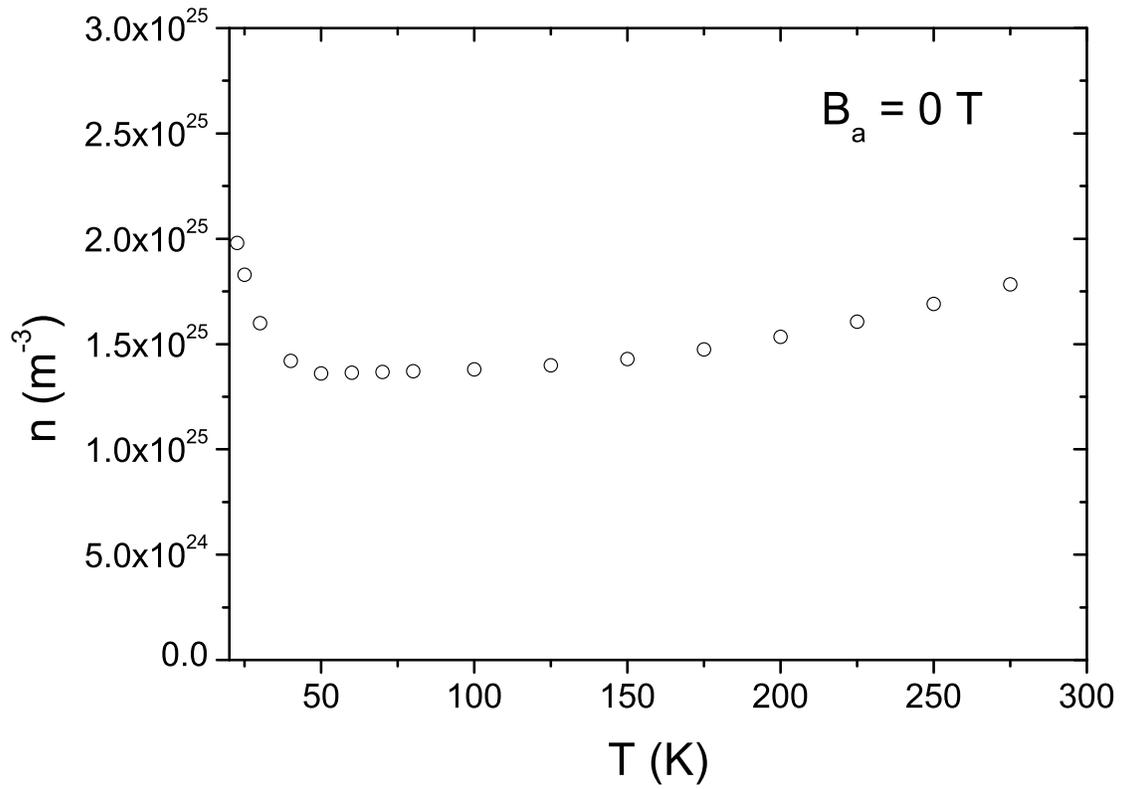}
  \caption{Itinerant carrier density $n(T)$ in EuB$_6$ at high temperatures, obtained from the separation
  of the total resistivity into a magnetic, a phononic and an impurity contribution. (see fig. 1)}\label{nhighT}
\end{figure}

In the next step we compare eq. (\ref{equation32}) with the
measured temperature-dependent resistivity. To begin with, we
postulate that the mobile charge carriers in the conduction band
originate from the transfer of electrons from doubly and singly
occupied levels of B$_6$-vacancies. The existence of such defects
has been invoked by Noack and Verhoeven \cite{noack} to explain
their gravimetric data on zone refined LaB$_6$. Their formation
energy has been shown to be substantially smaller than that of six
widely separated B-vacancies \cite{monnier}. An excellent fit is
obtained in the range 40 K $\leq$ $T$ $\leq$ 100 K with a constant
carrier concentration of 1.4$\cdot$10$^{25}$ m$^{-3}$ or 10$^{-3}$
/ unit cell, which corresponds to a Fermi energy $E_F$ of 54 meV.
At elevated temperatures, the experimental data suggest that
electrons from a narrow "band" of defect states which, for reasons
that are elucidated below, we associate with compensating ionized
acceptors in the form of Eu-vacancies, start to populate the
conduction band. The experimental data is well reproduced if we
assume a concentration of 6$\cdot$10$^{25}$ m$^{-3}$ defect
levels, with a lorentzian energy distribution centered at 19 meV
below the conduction band edge (i.e., 73 meV below $E_F$) and a
full width at half maximum of 9 meV. Finally, our fit requires the
density of mobile charge carriers to increase by 40 $\%$ as the
temperature is reduced from 40 K to 22.5 K. This increase can be
explained by an early onset of magnetic short range order
\cite{nyhus}, nucleated by the presence of defects, which locally
reduces the activation energy of the donor states. The different
components of the resistivity, their sum, and the measured curve
are displayed in Fig. \ref{rhovsT}. In contrast to earlier work
\cite{cooley}, where the contribution from electron-phonon
scattering to the room temperature (RT) resistivity was estimated
to be less than 3 $\%$, our analysis shows that this mechanism
actually dominates above 125 K and is responsible for over 60 $\%$
of the total resistivity at room temperature. Due to the small
size of $\rho_d$ $+$ $\rho_{cont}$, an unambiguous estimate of the
contact term is not possible at this stage and requires the
analysis of the magnetoresistance given below. The variation of
the charge carrier density with temperature is summarized in Fig.
\ref{nhighT}.

\subsubsection{Magnetoresistance}

\begin{figure}
  \centering
  \includegraphics[width=\linewidth]{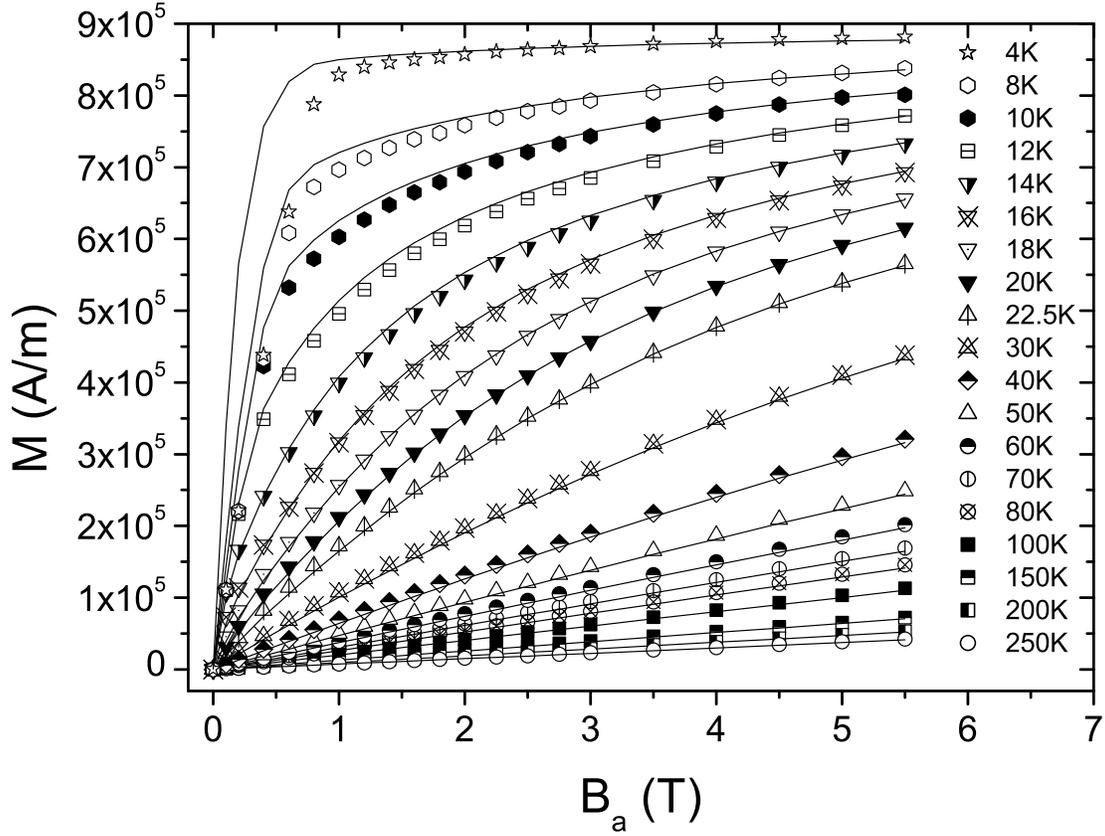}
  \caption{Bulk magnetization $M$ of EuB$_6$ as a function of applied magnetic
  field $B_a$ oriented perpendicularly to the platelet-shaped sample. All data
  for temperatures above 30 K, where fitted according to eq.
  (\ref{equation15}), yielding the parameters $M_{sat}$ and
  $\gamma$. The solid lines represent the mean field calculations
  for all temperatures using these parameters. Good agreement between this type of calculation and experiment prevails to even lower temperatures.}
\label{Magnetization}
\end{figure}

The magnetization of our sample as a function of the applied
magnetic field is displayed for a large number of temperatures in
Fig. \ref{Magnetization}, which also shows the results of a fit
using Eq. (\ref{equation15}) to all measured temperatures and
fields above 30 K. The latter yields a saturation magnetization of
$(8.83\pm0.04)\cdot 10^{5}$ A/m, in excellent agreement with the
value of $8.86\cdot10^{5}$ A/m expected for divalent europium, and
an effective molecular field parameter $\gamma = 5.15 \pm 0.05$.
Besides the Weiss field, $\gamma$ contains the Lorentz field
($\gamma_{L}=(1/3)$), negligible in higher $T_{C}$ materials, and
the demagnetizing field ($\gamma_{D} \approx -0.93$ for our
geometry). The Curie temperature of an infinite size bulk sample
is determined by the first two terms and amounts to 13.6 K, close
to the temperature at which neutron scattering experiments
\cite{heng} reveal the onset of spontaneous magnetic order.

The parameters also allow to calculate the longitudinal and
transverse susceptibilities using eqs. (\ref{equation18}) and
(\ref{equation17}), respectively, as well as the shift of the
bottoms of the spin-up and spin-down conduction bands, induced by
the non-zero magnetization, via eq. (\ref{equation5}). The latter
leads to a redistribution of charge carriers between the two bands
which, in turn, requires an adjustment of the chemical potential
with respect to the band minima. The Eu-vacancy levels will also,
to a lesser extent, be affected by the magnetization. The
spin-down states will rise in energy and progressively empty
themselves into the (spin-up) conduction band. At some
temperature-dependent value of the field, the latter will merge
with the spin-up Eu-vacancy states. Due to the Pauli principle,
the transport properties will not be affected, however.

The resistivity may now be calculated as follows. For all values
of the applied field and temperature, we define two average
relaxation rates

\begin{equation}\label{equation33}
   \frac{1}{\tau_{sd}^{\pm}}= \frac{1}{\sigma_{sd}^{\pm}(B_{a},T)}\frac{n^{\pm}e^{2}}{m^{*}}
   \tag{37}
\end{equation}
due to spin-disorder scattering. According to the model of Mandrus
et al. \cite{mand} the electron-phonon relaxation rate is
proportional to the Fermi velocity and hence, because $v_F$ $\sim$
$n^{1/3}$, we can write

\begin{equation}\label{equation34}
   \frac{1}{\tau_{ph}^{\pm}}=
   \left(\frac{n^{\pm}}{(n/2)}\right)^{1/3}
   \frac{1}{\tau_{ph}^{0}}= \frac{e^2}{m^*}n
   \left(\frac{n^{\pm}}{(n/2)}\right)^{1/3}\rho_{ph}(0,T) \ \ .
   \tag{38}
\end{equation}

Finally, we assume that the (weak) scattering by point defects is
field-independent, which leads to the total average relaxation
rate

\begin{equation}\label{equation35}
   \frac{1}{\overline{\tau}^{\pm}}=\frac{e^2}{m^*} \left[
   n^{\pm}\frac{1}{\sigma_{sd}^{\pm}(B_{a},T)}+n
   \left(\frac{2n^{\pm}}{n}\right)^{1/3}\rho_{ph}(0,T)+n\rho_{d}(0,T)\right]
\ \ ,
   \tag{39}
\end{equation}

and to the total resistivity in the presence of a magnetic field

\begin{equation}\label{equation36}
   \rho(B_{a},T) =\frac{m^*}{e^2}\left( n^{+}\overline{\tau}^{+}+
   n^{-}\overline{\tau}^{-}\right)^{-1}+\rho_{cont}
   \tag{40}
\end{equation}

In passing we note a spin-polarization of the itinerant electrons,
arising from a redistribution of the charge carriers between the
spin-up and spin-down bands, resulting from the opposite shifts of
the band edges $\epsilon_{0}^{+}$ and $\epsilon_{0}^{-}$ described
by eq. (\ref{equation5}). These shifts alone leave the density of
mobile charge carriers constant but the negative shift of the
majority band leads to a transfer of carriers from localized
defect to itinerant band states.

The free parameters in the model are the exchange coupling
constant $J$, the effective mass $m^*$, the contact resistivity
$\rho_{cont}$ and the charge carrier density $n_{tot}(B_a)$. The
best agreement with experiment is obtained for $J=0.18$ eV, $m^* =
0.22\cdot m_e$, where $m_e$ is the free-electron mass, and
$\rho_{cont}$ = 1.5$\cdot$10$^{-7}$ $\Omega$m. A ten percent
(correlated) variation of the parameters still produces reasonable
results. Our optimum value for the exchange coupling constant is
very close to the one quoted by Rys et al. \cite{balti} for
divalent europium (0.188 eV) and our value for $m^*$ compares well
with the density of states mass $m_{DOS}$ = 0.26$\cdot m_e$
yielded by the LSDA bandstructure calculation \cite{massidda}. The
absolute values of the carrier densities $n_{\pm}(B_a)$ depend
strongly on $J$ and $m^*$, but their relative changes are
identical for all parameters.

\begin{figure}
  \centering
  \includegraphics[width=\linewidth]{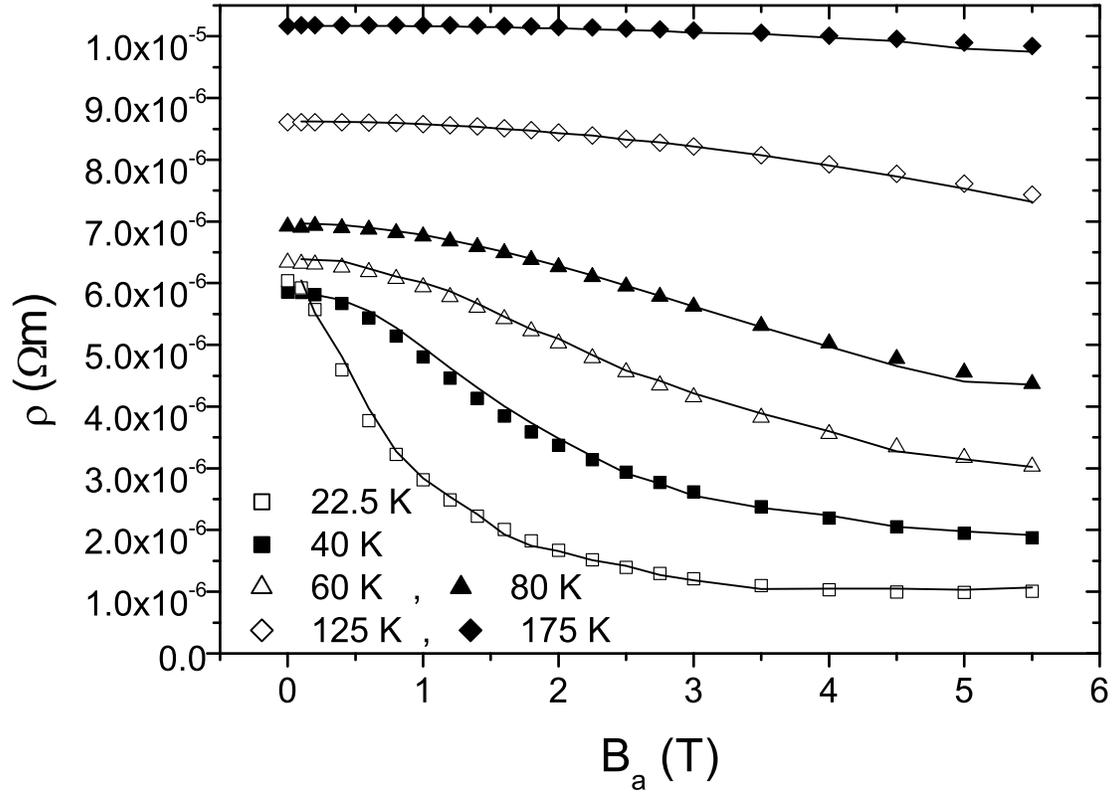}
  \caption{Magnetoresistivity of EuB$_6$ at 22.5, 40, 60, 80, 125 and 175 K, between 0 and 5.5 T. The solid lines are
  the results of the resistivity calculations described in the text.}
\label{magnetorhohighT}
\end{figure}

Figure \ref{magnetorhohighT} displays the measured curves for
$\rho(B_a,T)$ at 22.5, 40, 60, 80, 125 and 175 K. The solid lines
represent the calculations at the corresponding temperatures and
fields. We note a perfect agreement at temperatures above 60 K. At
22.5 K and fields less than  $\sim$ 2 T, strong polarization
effects induce a substantial variation of $n(B_a)$ which is
difficult to model. Nevertheless, the calculated curve reproduces
the measured results to within 5 $\%$. For stronger magnetic
fields the measured curve for $\rho(B_a)$ decreases more slowly
with increasing magnetic field, reflecting a further reduction of
the spin-disorder scattering and an increase of the charge carrier
density. Eventually, $\rho(B_a)$ flattens out and subsequently
increases slightly towards the highest fields, which we interpret
as the onset of conductivity through a second band, described in
more detail in the section on the low-temperature transport. The
charge carriers in this second band are holes with a concentration
increasing from 0 at 4 T to $n_h$ = 0.8$\cdot$10$^{25}$m$^{-3}$ at
5.5 T.

\begin{figure}
  \centering
  \includegraphics[width=\linewidth]{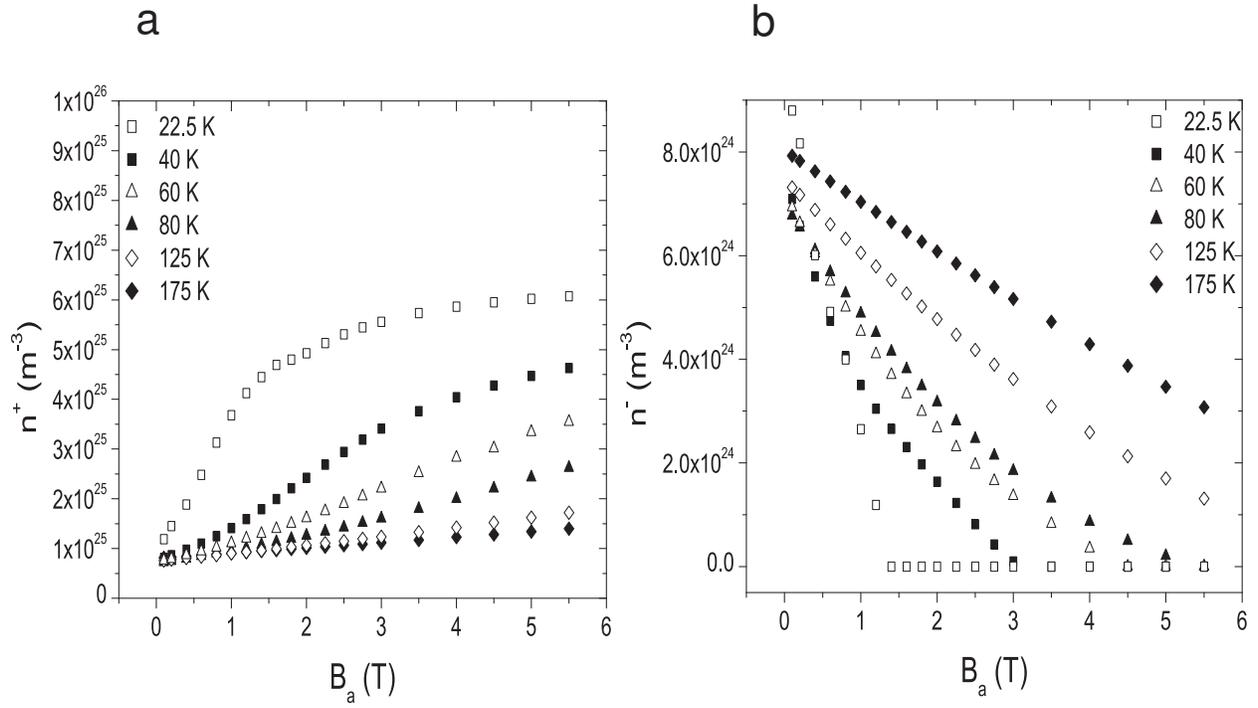}
  \caption{The charge carrier concentrations $n^+(B_a)$ (a) and $n^-(B_a)$ (b) in the spin moment up and spin moment down
band, respectively,
  at 22.5, 40, 60, 80, 125 and 175 K between 0
  and 5.5 T. Note the different scales of the y-axis.}
\label{nplus}
\end{figure}

In Figures \ref{nplus} a and b we display the electron densities
$n^+$ and $n^-$ in the spin-up and spin-down band, respectively,
for the same temperatures and fields for which $\rho(B_a)$ was
calculated. With decreasing temperature, the polarization effects
lead to a stronger enhancement of $n^+$ and a corresponding
reduction of $n^-$ with increasing field. At 60 K $n^{-}$ vanishes
at $\approx$ 4.5 T, leaving a fully polarized conduction band. At
40 K, $n^{-}$ vanishes at 3 T and at 22.5 K already at 1.4 T. As
mentioned before, an increasing concentration of holes has to be
introduced below 40 K in order to explain the high field ($B_a >$
4 T) data. The only plausible mechanism for this to happen is
that, e.g., at 22.5 K and 4 T, the top of the spin moment up
valence band which, according to the calculation of Massidda et
al. \cite{massidda} should experience an (upward) shift of the
order of 15 percent of that of the bottom of the conduction band,
touches the Fermi level $E_F$. Note that this is not in
contradiction with the existence of ionized (i.e., occupied by
electrons) acceptor states below $E_F$.

\subsubsection{Hall effect}

In a magnetic conductor, the Hall resistance consists of two
contributions, namely the ordinary part $\rho_{H}^{ord}$, due to
the Lorentz force $e \overrightarrow{v}\times \overrightarrow{B}$
acting on the electrons, and the anomalous part $\rho_{H}^{mag}$,
which results from the antisymmetric scattering of itinerant
charge carriers by the disordered local moments on the Eu-ions
\cite{Hirsch}. The spin-flip exchange scattering mixes the states
of the spin-up and spin-down conduction bands, which can therefore
be considered as a single entity. Therefore the ordinary Hall
resistivity is related to the total density of mobile charge
carriers, $n_{tot}$ by the usual relation
\begin{equation}\label{equation37}
  \rho_{H}^{ord} = -\frac{1}{n_{tot}e}\cdot B \ \ \ ,
  \tag{41}
\end{equation}

with $\overrightarrow{B}= \overrightarrow{B_{a}}+
  \mu_{0}(1-\gamma_{D})\overrightarrow{M}$.

\begin{figure}
\centering
  \includegraphics[width=\linewidth]{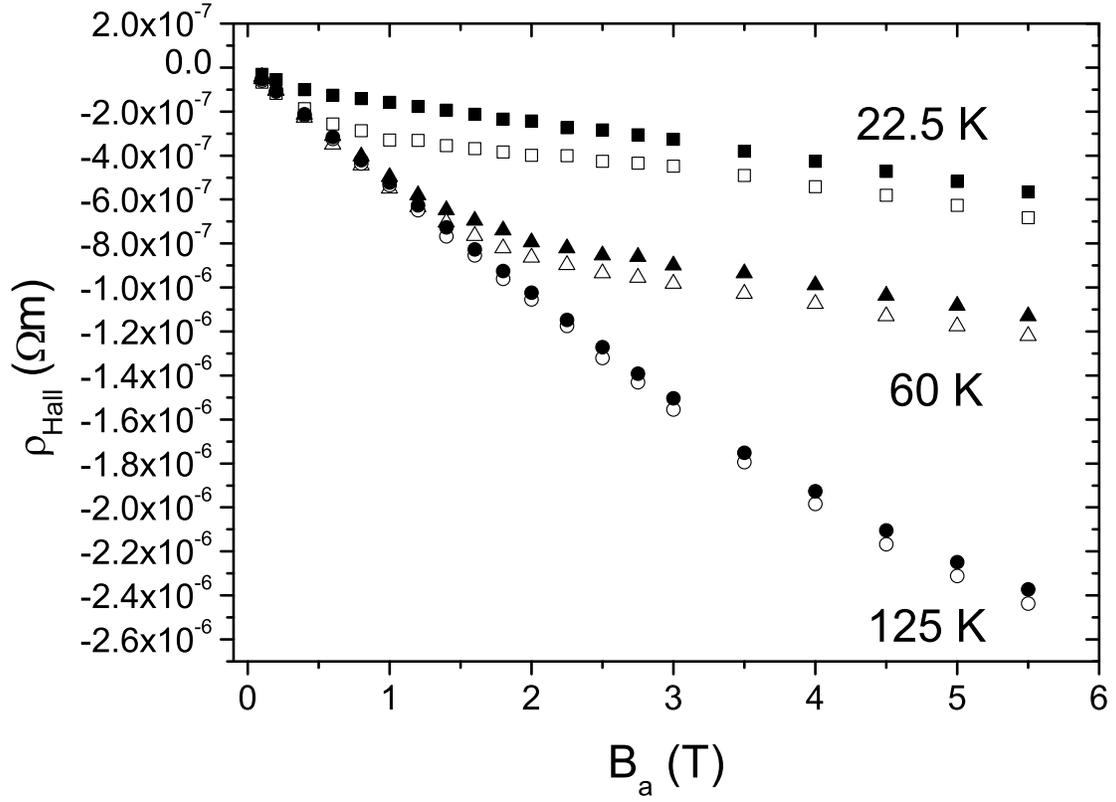}
  \caption{Measured Hall resistivities $\rho_{Hall}(B_a)$ and calculated ordinary Hall resistivities $\rho_H^{ord}$
    of EuB$_6$ at 22.5, 60 and 125 K between 0 and 5.5 T. The
empty symbols show $\rho_{Hall}(B_a)$, the full symbols display
$\rho_H^{ord}$.}
 \label{RHallord}
\end{figure}

In Fig. \ref{RHallord} we display the measured Hall resistivity
$\rho_H$ as a function of applied field for 22.5, 60 and 125 K,
together with $\rho_H^{ord}$ computed with the charge carrier
densities obtained from the fit to the magnetoresistivity data.
The difference between the measured Hall resistivity and
$\rho_H^{ord}$ is largest at low temperatures and in small fields,
where the spin up and spin down carrier densities are most
sensitive to fluctuations in the magnetization (see Fig.
\ref{nplus}).

\begin{figure}
  \centering
  \includegraphics[width=\linewidth]{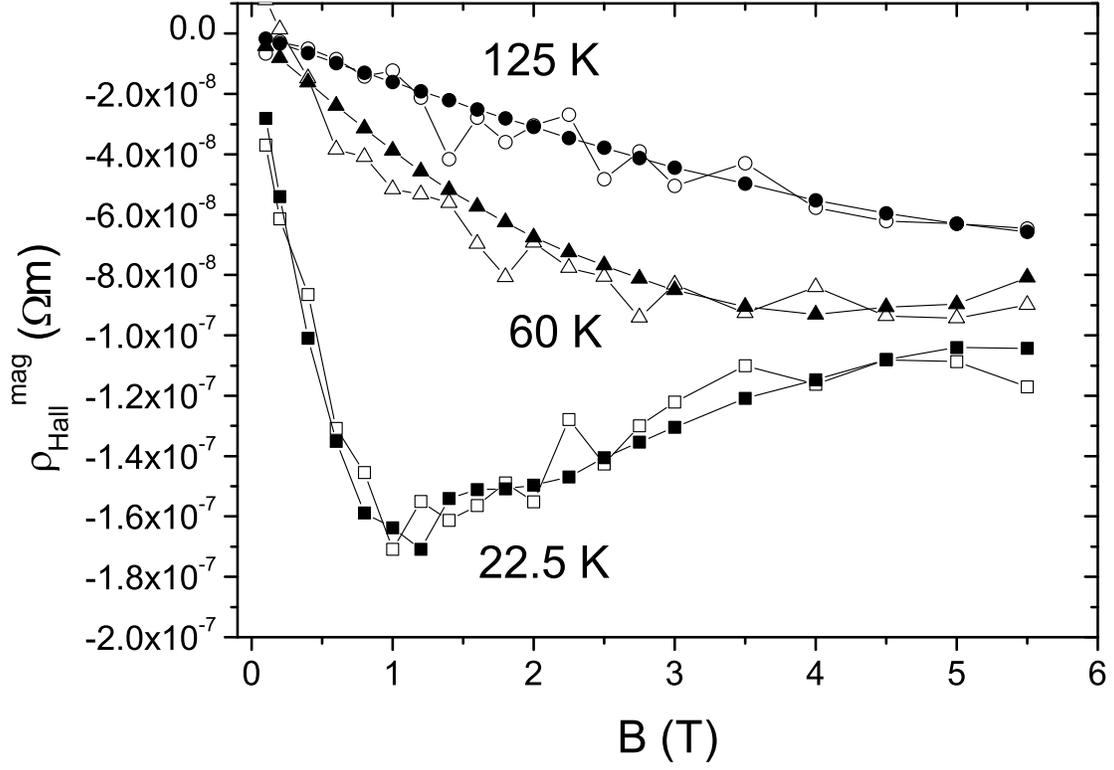}
  \caption{Calculated and experimentally derived anomalous Hall resistivities $\rho_{H}^{mag}(B_a)$ of EuB$_6$ at 22.5, 60 and 125 K between 0
  and 5.5 T. The empty symbols show the difference between the measured
  Hall resistivity $\rho_{H}$ and the corresponding calculated ordinary contribution
  $\rho_{H}^{ord}$.
  The full symbols display the calculated anomalous Hall
  resistivities $\rho_{H}^{mag}$.}
\label{Hallmag}
\end{figure}

We attempted to model this difference, which we interpret as the
anomalous Hall resistivity $\rho_H^{mag}$, using the relaxation
rate for antisymmetric scattering given by equation (\ref{H10}).
Treating $\widetilde{C}$ as a free parameter in a fit to
$\rho_H^{mag}$ with $\rho_H^{mag}$ =
$(\sigma_H^{mag,+}+\sigma_H^{mag,-})^{-1}$, and inserting the Hall
conductivities obtained from eq. (\ref{equation28}) with
$\tau_{sd}$ replaced by $\tau_{as}$, yields the curves shown in
Fig. \ref{Hallmag} and the optimum value $\widetilde{C}$ $\approx$
(6.6 $\pm$ 0.5)$\cdot$10$^{-11}$ m, exceeding that of
$\widetilde{C}^{M}$ (eq. (\ref{H12})) by more than four orders of
magnitude. Using the value of $m^{*}$ obtained from our fit to the
magnetoresistivity, the spin-orbit coupling constant $\lambda$ =
164 meV for the intermediate 4f$^{6}$ configuration \cite{so} and
the lattice constant a = 4.185 \AA, we can write the corresponding
coefficient for the Kondo mechanism as

\begin{equation}\label{H15}
   \widetilde{C}^{K} = 1.77\cdot
   10^{-13}\frac{V_1^2}{\Delta_{-}^{2}} \ \textrm{m} \ \ .
\tag{42}
\end{equation}

For $\widetilde{C}^{K}$ to adopt the optimum value obtained from
the fit to $\rho_H$ $-$ $\rho_H^{ord}$ would require the ratio
$V_1/\Delta_{-}$ to be of the order of 20, which is utterly
unrealistic. According to X-ray photoemission experiments
\cite{xray}, the lower bound on $\Delta_{-}$ ($^{7}$F$_0$ final
state) is $\sim$ 0.7 eV and $V_1$ is expected to be smaller. It
appears that discrepancies of that order are the rule for systems
with half-filled 4f-shells in their ground state, such as
gadolinium and Gd compounds \cite{kondo, christen}.

\section{Low-Temperature Transport}\label{lowTT}

\begin{figure}
  \centering
  \includegraphics[width=\linewidth]{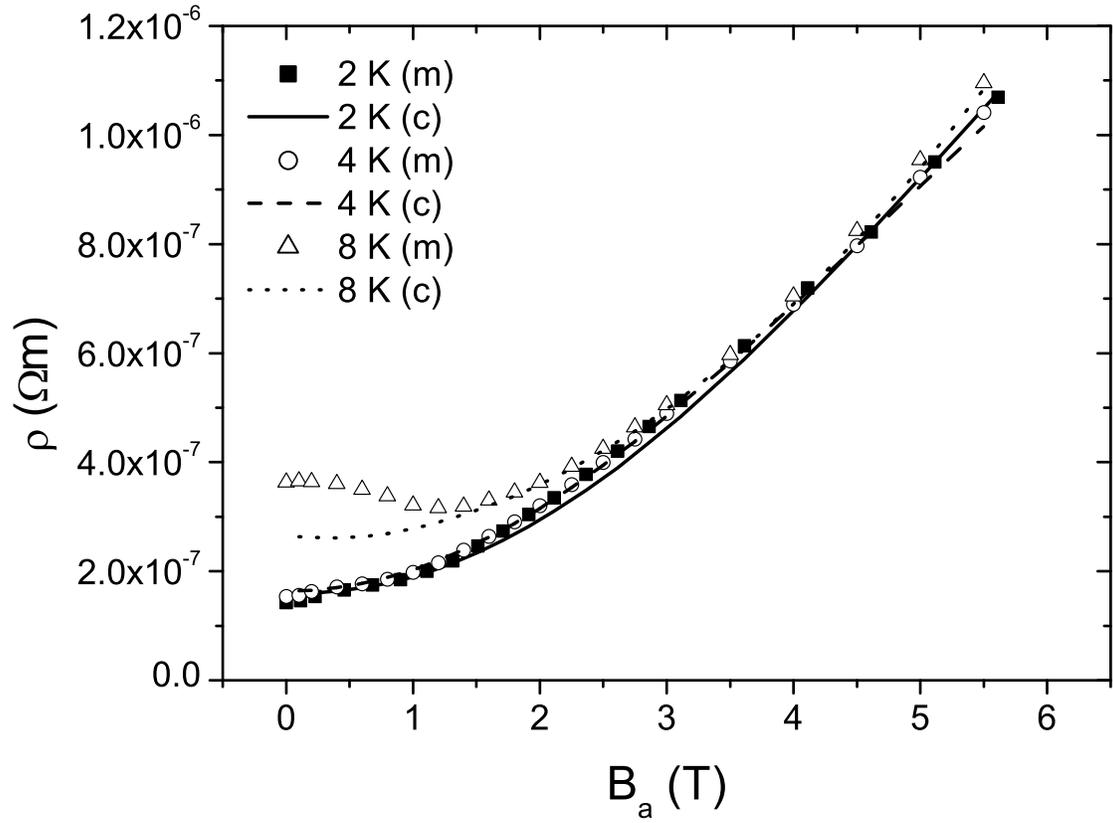}
  \caption{The measured (m) magnetoresistivity data $\rho(B_a)$ for EuB$_6$ are displayed for 2, 4 and 8 K between 0 and 5.5
T. The calculated (c) curves are obtained for the corresponding
temperatures, using the two-band model captured by eqs.
(\ref{equation39a}) and (\ref{equation39b}).}

\label{spinwaverho}
\end{figure}

\begin{figure}
  \centering
  \includegraphics[width=\linewidth]{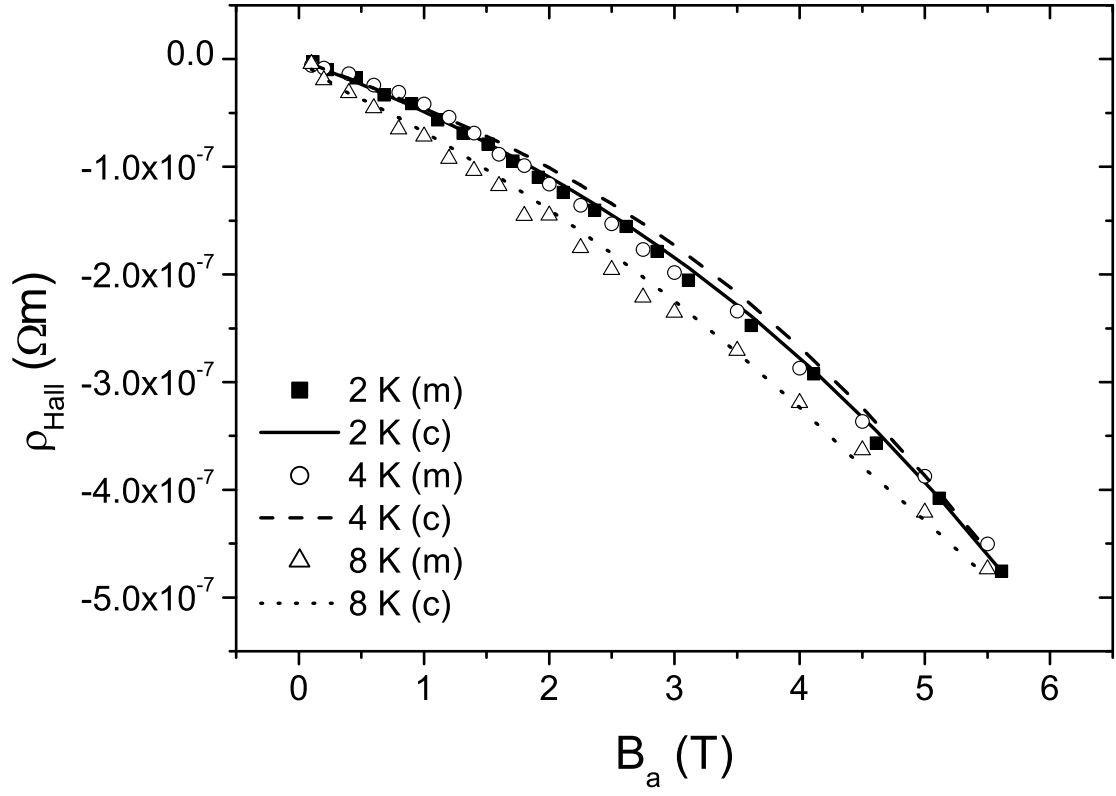}
  \caption{The measured (m) Hall resistivity data $\rho_H(B_a)$ for EuB$_6$ are displayed for 2, 4 and 8 K between 0 and 5.5
T. The calculated (c) curves are again obtained, using the
two-band model captured by eqs. (\ref{equation39a}) and
(\ref{equation39b}).}
  \label{spinwaverhoHall}
\end{figure}

At temperatures below 8 K, electron-phonon scattering is
negligibly small, and the only contributions to the resistivity
are $\rho_{sd}$, $\rho_d$ and $\rho_{cont}$. In figures
(\ref{spinwaverho}) and (\ref{spinwaverhoHall}), the data for the
magnetoresistance and the Hall resistance are plotted for 2, 4 and
8 K. At 2 K, $\rho(B_a)$ increases by a factor of $\approx$ 7
between 0 and 5.5 T. In fields exceeding 1.5 T, $\rho (B_a)$ is
nearly quadratic in $B_{a}$ for all three temperatures. This
observation strongly suggests that two bands with oppositely
charged carriers participate in the conduction of electrical
current. For $B_a$ $>$ 1.5 T, we therefore use a standard two-band
model \cite{Ashcroft} to simultaneously describe $\rho_{H}(B_{a})$
and $\rho(B_{a})$. This leads to

\begin{equation}\label{equation39a}
   R = \frac{R_{1} \rho_{2}^{2}+R_{2} \rho_{1}^{2}+R_{1}
   R_{2}(R_{1}+R_{2})B_{a}^{2}}{(\rho_{1}+\rho_{2})^{2}+(R_{1}+R_{2})^{2}B_{a}^{2}}
\tag{43a}
\end{equation}

and

\begin{equation}\label{equation39b}
   \rho = \frac{\rho_{1} \rho_{2}(\rho_{1}+\rho_{2})+(\rho_{1} R_{2}^{2}+\rho_{2}
   R_{1}^{2})B_{a}^{2}}{(\rho_{1}+\rho_{2})^{2}+(R_{1}+R_{2})^{2}B_{a}^{2}}
   + \rho_{cont}
\ \ , \tag{43b}
\end{equation}

where $R_1$, $R_2$, $\rho_2$ and $\rho_2$ depend on temperature
and on the applied field. The Hall "constant" $R$ is the
proportionality factor between $\rho_{H}(B_{a})$ and $B_{a}$;
$R_{1}$ and $R_{2}$ are the Hall "constants" for the conduction
and the valence band, respectively.

At 8 K and in zero external field, the ordered Eu moment is of
equal magnitude \cite{heng} as the net moment per Eu ion in the
field direction at 22.5 K and $B_a$ = 5.5 T. Hence we expect to
find the same concentration of electrons $n_e$ $\approx$
6.1$\cdot$10$^{25}$ m$^{-3}$ in the (fully polarized) conduction
band and the same value of $R_1$(8 K, 0 T) $\approx$ $-$
1.0$\cdot$10$^{-7}$ m$^3$A$^{-1}$s$^{-1}$ in both cases. Similarly
$n_h$ $\approx$ 0.8$\cdot$10$^{25}$ m$^{-3}$ and $R_2$(8 K, 0 T)
$\approx$ 7.7$\cdot$10$^{-7}$ m$^3$A$^{-1}$s$^{-1}$. We determine
$\rho_1$ and $\rho_2$ under the same conditions as follows. First
we note (Fig. (\ref{spinwaverho})) that the contribution from
spin-disorder scattering to the total resistivity is negligibly
small for $B_a$ $\geq$ 1.5 T. For applied fields in excess of this
value $\rho_1$ ($\rho_2$) is therefore entirely due to the
scattering of electrons (holes) by point defects, and is
proportional to the density of electrons (holes), with no explicit
dependence on $B_a$. This allows us to extrapolate this
contribution, which we call $\rho_{1d}$ ($\rho_{2d}$), to zero
applied field as follows

\begin{equation}\label{backcalc}
  \rho_{1d}(8K,0T) = \rho_{1d}(22.5K,5.5T)=
  \frac{n_{e}(22.5K,0T)}{n_{e}(22.5K,5.5T)}\cdot\rho_{1d}(22.5K,0T)\approx
  0.66\cdot 10^{-7} \Omega m
\tag{44}
\end{equation}

where the carrier densities can be read off Fig. (\ref{nplus}),
and $\rho_{1d}$(22.5 K, 0 T) is obtained from Fig. (\ref{rhovsT})
and $\rho_{cont}$ determined in the previous section. From the two
band model and in the absence of spin-disorder scattering it
follows that $\rho_{2d}$(8 K, 0 T) $\approx$ 1.44$\cdot$10$^{-6}$
$\Omega$m. Considering the ratio of the effective masses and the
carrier concentrations for the two bands, we find that the
relaxation time of the holes is approximately three times shorter
than that of the electrons. Enhancing the applied field from 0 to
5.5 T induces a monotonous enhancement of the ordered Eu moment
and thus the magnetization. This in turn enhances the overlap
between the valence band and the conduction as well as the donor
spin-up bands, leading to a net increase in the density of mobile
charge carriers. From the fit of our experimental data to eqs.
(\ref{equation39a}) and (\ref{equation39b}) for $B_a$ $\geq$ 1.5 T
we obtain $n_e$ = 6.7$\cdot$10$^{25}$ m$^{-3}$ and $n_h$ =
6.1$\cdot$10$^{25}$ m$^{-3}$ at magnetic saturation. The growth
rate is roughly proportional to $\left(M/M_{sat}\right)^{3/2}$, as
expected for parabolic bands. The residual spin disorder
resistivity in zero field amounts to less than 1$\cdot$10$^{-7}$
$\Omega$m (see Fig. (\ref{spinwaverho})).

Below 5 K the elementary excitations of the system of magnetic Eu
ions are spin-waves \cite{heng}. From a comparison of the
zero-field resistivities at 2 and 4 K we see that the scattering
of the charge carriers by these collective modes, which should be
proportional to $T^2$, can be neglected. The field dependence of
the (Hall) resistivity at these two temperatures is again well
reproduced by the two-band model, on which we have imposed the
constraint that the resulting values for $R_{1,2}$ and
$\rho_{1,2}$ at full magnetization are the same as at 8 K.

\section{Discussion}

\begin{figure}
  \centering
  \includegraphics[width=\linewidth]{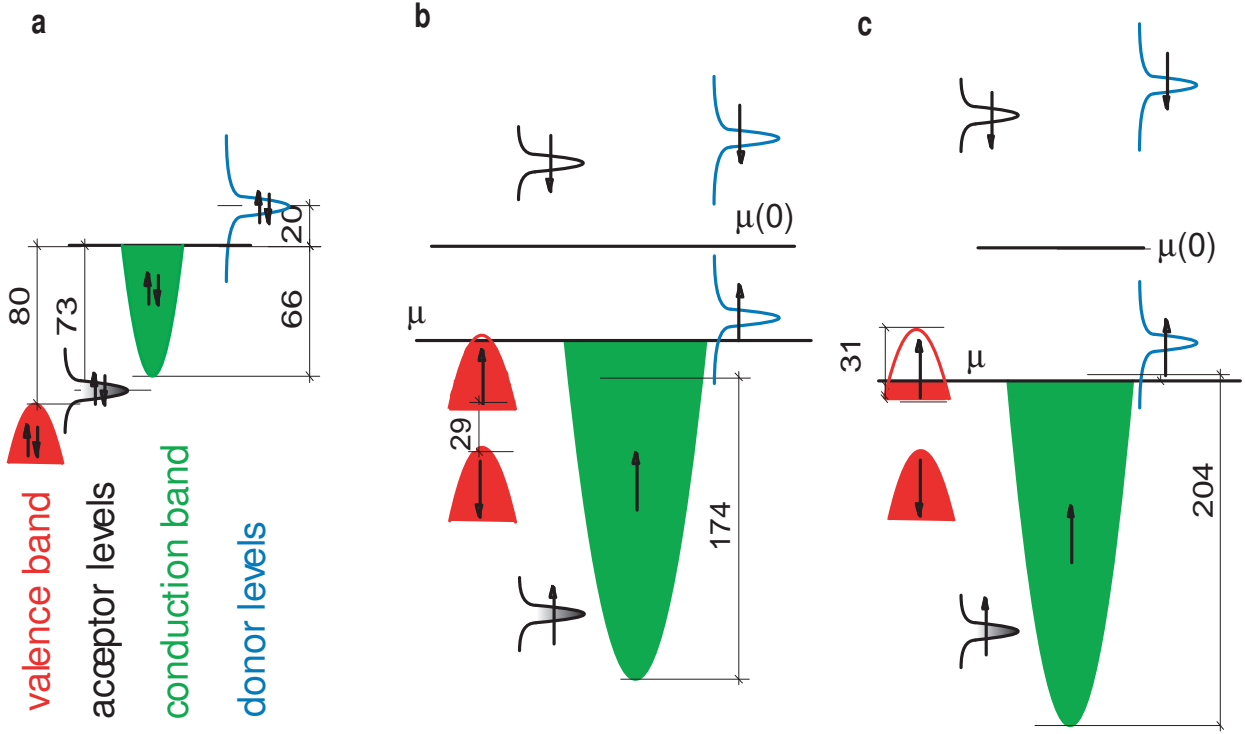}
  \caption{(Color in online edition) Schematic electronic excitation spectrum of EuB$_6$ around the chemical potential $\mu$ at $T$ = 22.5 K. A possible arrangement
of the conduction band, the valence band, an acceptor and a donor
defect band is plotted for a: $B_a$ = 0 T, b: $B_a$ = 4 T, c:
$B_a$ = 5.5 T. From left to right in each panel: valence band,
acceptor levels (cation vacancies), conduction band, donor levels
(B$_6$ vacancies). All energies are given in meV. The up-arrows
and down-arrows denote the spin moment up and spin moment down
subbands, respectively, $\mu(0)$ is the chemical potential for
$B_a$ = 0 T, whereas $\mu$ denotes the chemical potential at the
corresponding fields. The distribution of the charge carriers over
the 4 different bands is explained in the text. Note that in b)
and c) the conduction subband for the down moment lies far above
the chemical potential and is thus irrelevant for our purposes.}
 \label{bandstructure}
\end{figure}

In this paper we offer a consistent, quantitative description of
the magnetoresistance and the Hall effect in EuB$_6$ over a wide
range of temperatures above and below the magnetic phase
transition. From our analysis, the following picture of the
electronic structure of this compound emerges. For $T$ $>$ 20 K
and no applied magnetic field, EuB$_6$ is a heavily (self-) doped,
strongly compensated n-type semiconductor. The donors are B$_6$
vacancies, whose energy levels form a narrow "band" centered above
the chemical potential, and possibly trivalent impurities, which
we have neglected. The acceptors are cation vacancies, always
present in the hexaborides, whose energy levels also form a narrow
"band" just above the top of the valence band (Fig.
\ref{bandstructure} a). At 22.5K, the intrinsic band gap is of the
order of 14 meV and the chemical potential lies $\sim$ 66 meV
above the bottom of the conduction band. Upon application of a
magnetic field, the Eu-ions acquire a net moment in the field
direction. The latter couples to the conduction electrons through
the exchange Hamiltonian $\mathcal{H}_1$, leading to a splitting
of the conduction band proportional to the magnetization in first
order. The acceptor levels, the wave functions of which are mainly
composed of d$_{z^2}$ orbitals reaching out from the neighboring
cations, suffer a splitting of similar size, while that of the
donor levels and especially that of the valence band, which both
couple to the Eu moments only indirectly through hybridization
with the cation's d-orbitals, will be smaller \footnote{The
splitting of the acceptor levels is taken to be 75 $\%$, that of
the donor levels 50 $\%$ and that of the valence band 15 $\%$ of
the splitting of the conduction band, respectively.}. These
splittings lead to a redistribution of the charge carriers between
the different bands and, in our model, at a temperature dependent
critical value of the applied field, e.g., 1.4 T at 22.5 K, the
spin-down conduction band has emptied itself completely. At that
point, the magnetization has reached only 25 percent of its
saturation value. As the applied field and the magnetization are
further enhanced, the top of the spin-down valence band moves
closer to the selfconsistently determined chemical potential (Fig.
(\ref{bandstructure} b)), until it crosses it. At 5.5 T
($M/M_{sat}$ $\approx$ 0.65) the situation is that of Fig.
(\ref{bandstructure} c). Once the saturation magnetization has
been reached, the carrier densities in the valence and the
conduction band stay constant.

This picture is consistent with the observation of two ellipsoidal
pockets in de Haas van Alphen (dHvA) and Shubnikov de Haas
experiments, performed at fields above 5 T \cite{goodrich,aron}.
The electrons and holes being in different spin states, their
Bloch functions cannot mix. It also offers a natural explanation
for the weak temperature dependence of the dHvA frequencies, even
accross $T_C$ \cite{aron}, since these are only affected by
deviations of the 4f-electron based magnetization from its
saturation value. In view of the sensitivity of the system to
defects, our carrier concentrations are in very reasonable
agreement with the ones quoted in refs. \cite{aron} and
\cite{goodrich}.

\begin{figure}
  \centering
  \includegraphics[width=\linewidth]{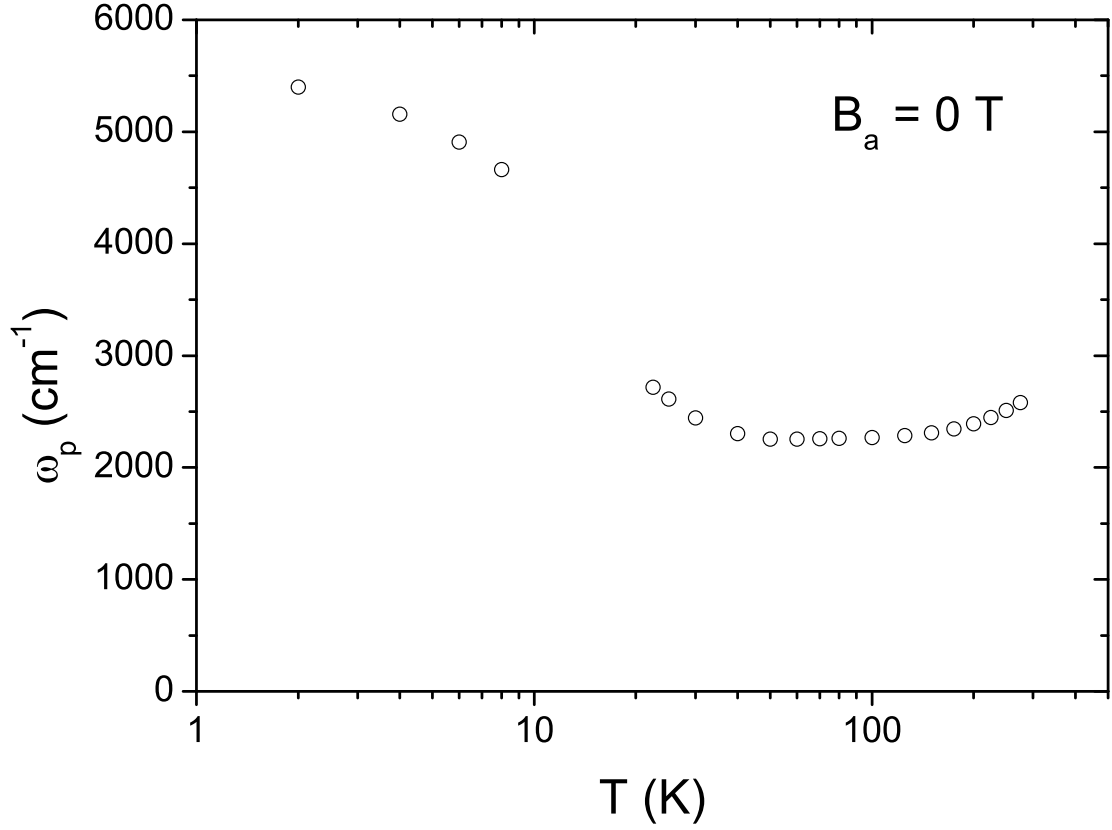}
  \caption{Bare plasma frequency $\omega_p$ of EuB$_6$, obtained from the calculated itinerant charge carrier densities, vs. temperature in zero magnetic field. $\omega_p$ is calculated using eq. (\ref{equation40}) and the optical massed provided by the band structure calculations ($m_e^{opt}$ = 0.24$m_e$, $m_h^{opt}$ =
0.29$m_e$). } \label{chargecarriers}
\end{figure}

A further test of our interpretation is provided by the
reflectivity experiments of Degiorgi and collaborators
\cite{leo,Broderick}. In Fig. \ref{chargecarriers} we display the
bare plasma frequency
\begin{equation}\label{equation40}
  \omega_p =
\left[\frac{e^2}{\epsilon_0}\left(\frac{n_e}{m_e^{opt}}+\frac{n_h}{m_h^{opt}}\right)\right]^{1/2}
\tag{45}
\end{equation}
as a function of temperature in zero field, calculated with the
carrier densities obtained from our fits and the optical masses
provided by the band structure calculations of ref.
\cite{massidda} ($m_e^{opt}$ = 0.24$m_e$, $m_h^{opt}$ =
0.29$m_e$). Our results compare well with the data of ref.
\cite{leo}. In particular, we reproduce, even quantitatively, the
steep rise of $\omega_p$ below $T_C$. The decrease of the computed
plasma frequency at temperatures between 300 K and 100 K is
consistent with the observed red shift of the plasma edge in ref.
\cite{leo}.

\begin{figure}
  \centering
  \includegraphics[width=\linewidth]{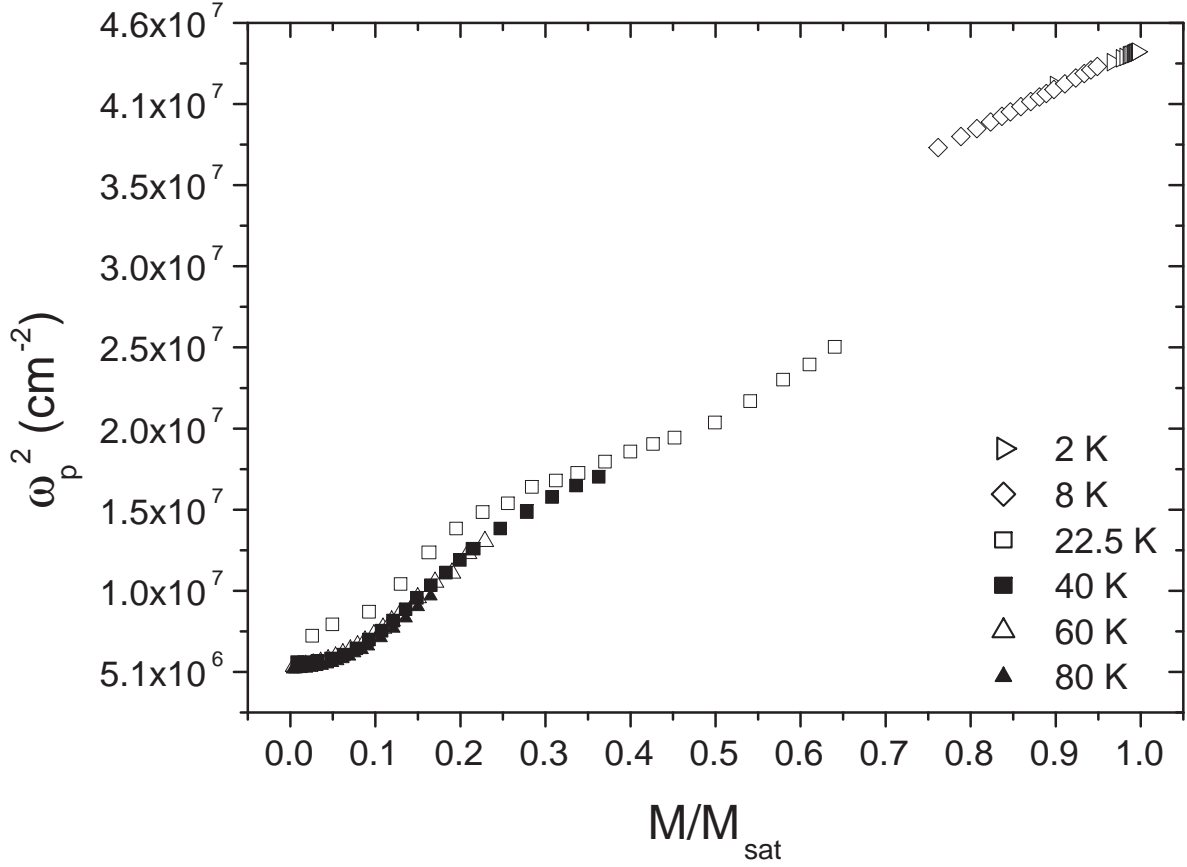}
  \caption{Squared bare plasma frequency $\omega_p^2$ of EuB$_6$ versus the relative bulk magnetization
  $M/M_{sat}$ at 2, 8, 22.5, 40, 60 and 80 K. $\omega_p$ is calculated using eq. (\ref{equation40}) and the optical massed provided by the band structure calculations ($m_e^{opt}$ = 0.24$m_e$, $m_h^{opt}$ =
0.29$m_e$).} \label{ntot}
\end{figure}

Fig. \ref{ntot} shows the dependence of the plasma frequency on
magnetization obtained from our model. In ref. \cite{Broderick},
$\omega_p^2$ was found to be proportional to $M$ for 1.6 K $\leq$
$T$ $\leq$ 35 K and 0 T $\leq$ $B_a$ $\leq$ 7 T. Again, our
results are compatible with this behavior, but suggest that the
relation between the two quantities may be more complex.

Finally we compare our model with the ARPES and bulk-sensitive XAS
and SXE data of ref. \cite{denlinger}, which were obtained in the
temperature range between 20 and 30 K. The existence of an X-point
electron pocket was assumed a priori in our theoretical ansatz,
relying on the validity of these experimental results. The authors
of ref. \cite{denlinger} attribute the feature labelled "band 1"
in their paper to the emission from the valence band. The fact
that its dispersion is much weaker in EuB$_6$ than in CaB$_6$ and
SrB$_6$, and the value of its binding energy of $\sim$ 1.2 eV,
lead us to interpret it as an emission from the Eu 4f-shell.
According to our model, the emission from the valence band should
start at a binding energy of $\sim$ 0.1 eV, which is not seen in
ARPES. This may be due to the fact that the exposed $[$100$]$ face
consists of metal atoms only and that the electrons, originating
from the boron network, cannot escape from the solid at the given
photon energies. Another complication is the observed
time-dependent surface relaxation \cite{denlinger}. The SXE and
XAS data are consistent with our interpretation of the ARPES data.

\section{Conclusion}

Although "real" EuB$_6$ is a heavily doped, strongly compensated,
and therefore very disordered magnetic semiconductor, many of its
properties can be satisfactorily described by a relatively simple
model, taking into account the two main intrinsic sources of
imperfections, namely Eu and B$_6$ vacancies. Our microscopic
treatment above $T_C$ is limited to the range of temperatures
where mean field theory can be used to describe the magnetic
properties of the compound. We expect the behavior around $T_C$
(and the critical temperature itself) to be sample dependent, as
the (RKKY) coupling between magnetic ions is mediated by the
conduction electrons whose concentration is a function of the
magnetization.

What is still missing is a plausible mechanism leading to the
correct order of magnitude for the anomalous Hall resistivity in
this and the other compounds formed by rare-earth ions with a
half-filled 4f shell. In our opinion the order of magnitude
discrepancy between theory and experiment with respect to the
anomalous Hall effect is not caused by underestimating the mixing
matrix element $V_1$, as this would reflect itself in the
resistivity as well. It must therefore be connected with the
multiplicity of the intermediate states which can be coupled by
the spin-orbit interaction. The lowest energy term for the 4f$^6$
configuration is characterized by $L$ = 3 and $S$ = 3, with a
degeneracy of 49 in the absence of spin-orbit coupling. Whereas
the unit operator, relevant for the resistivity, only has diagonal
matrix elements, the spin-orbit operator, which splits the term
into seven multiplet levels $J$ = 0, 1,... 6, has matrix elements
between states satisfying the selection rule $\Delta J_z$ = 0,
$\pm$1, within every subspace corresponding to a given value of
$J$. This leads to 133 possible transitions instead of 49, still
not enough to account for the observed difference. We conjecture
that the small hybridization between the europium (gadolinium) 4f
orbitals and the boron sp \cite{massidda, hasegawa} (gadolinium 5p
\cite{temm}) orbitals, suggested but overestimated by band
structure calculations, opens the necessary extra channels. We
hope that our results, which confirm previously established
discrepancies, will encourage more theoretical work on this long
standing problem.

\section{Acknowledgments}
We thank M. Chiao for letting us use her resistivity data on
YbB$_6$ prior to publication. Stimulating discussions with her and
L. Degiorgi are gratefully acknowledged. This work has benefitted
from partial financial support of the Schweizerische Nationalfonds
zur F\"{o}rderung der wissenschaftlichen Forschung and the US-NSF
grant DMR-0203214.

\newpage

\end{document}